\documentclass[12pt]{article}

\usepackage{amsmath}
\usepackage{graphicx}
\usepackage{mathptmx}
\usepackage{amssymb,amsfonts,amsthm,bm}
\usepackage[ansinew]{inputenc}
\usepackage{helvet,times}
\usepackage{bm,textcomp}
\usepackage{ulem}

\usepackage[margin=1in]{geometry}

\usepackage{natbib}
\setcitestyle{round,authoryear,semicolon,aysep={,}}

\usepackage{url}
\makeatletter
\renewcommand\@biblabel[1]{#1.}
\makeatother

\begin{document}

\title{
    Inversion of hematocrit partition at microfluidic bifurcations
}
\author{
    Zaiyi Shen,$^1$ Gwennou Coupier,$^1$ Badr Kaoui,$^{2,3}$ Beno\^{i}t Polack,$^{4,5}$ \\Jens Harting,$^{6,7}$ Chaouqi Misbah,$^1$ Thomas Podgorski$^1$$^\ast$
}
\date{\today}

\maketitle
\begin{description}
    \item[affiliations:]
$^1$ Laboratoire Interdisciplinaire de Physique (LIPhy) UMR5588 CNRS-Universit\'e Grenoble Alpes, Grenoble F-38041, France ; 
$^2$ CNRS-Sorbonne University, Universit\'e de Technologie de Compi\`egne, UMR7338 Biomechanics and Bioengineering, 60203 Compi\`egne, France;
$^3$ TP1, University of Bayreuth, D-95447 Bayreuth, Germany; 
$^4$ Laboratoire d'H\'ematologie, CHU, Grenoble, France; 
$^5$TIMC-IMAG/TheREx, CNRS UMR5525, Universit\'e Grenoble Alpes, Grenoble, France; 
$^6$ Department of Applied Physics, Eindhoven University of Technology, P.O. Box 513, 5600MB Eindhoven, The Netherlands; 
$^7$ Faculty of Science and Technology, Mesa+ Institute, University of Twente, 7500 AE Enschede, The Netherlands
    \item[abbreviated title:]
        Inversion of hematocrit partition
    \item[corresponding author:]
        Thomas Podgorski,
        e-mail: thomas.podgorski@ujf-grenoble.fr
\end{description}


\begin{abstract}
Partitioning of red blood cells (RBCs) at the level of bifurcations in the microcirculatory system affects many physiological functions yet  it remains poorly understood. We address this problem by using T-shaped microfluidic bifurcations as a model. Our computer simulations and \textit{in vitro} experiments reveal that the hematocrit ($\phi_0$) partition depends strongly on RBC deformability, as long as $\phi_0 <20$\% (within the normal range in microcirculation), and can even lead to complete deprivation of RBCs in a child branch. Furthermore, we discover a deviation from the Zweifach-Fung effect which states that the child branch with lower flow rate recruits less RBCs than the higher flow rate child branch. At small enough $\phi_0$, we get the inverse scenario, and the hematocrit in the lower flow rate child branch is even higher than in the parent vessel. We explain this result  by an intricate up-stream RBC organization and we highlight the extreme dependence of RBC transport on geometrical and cell mechanical properties. These parameters can lead to unexpected behaviors with consequences on the microcirculatory function and oxygen delivery in healthy and pathological conditions.
 
\end{abstract}

\textbf{keywords:} microcirculation; blood; red blood cell; microfluidics; lattice Boltzmann method


\section{Introduction}
\label{sec:introduction}

Blood flows through a complex network of the circulatory system -- from large arteries to very tiny capillaries -- in order to ensure oxygen delivery and to remove metabolic waste. This task is mainly carried out by red blood cells (RBCs) that are remarkably deformable, in healthy conditions, and therefore able to squeeze into tiny capillaries. A change in rheological and flow properties of the blood is often associated with hematological diseases or disorders \citep{fedosov11} (e.g. sickle-cell anemia, malaria, polycythemia 
vera). Understanding blood flow and its dependence on the mechanical properties of its constituents may improve and lead to new applications in biomedical technology, for example in blood substitutes development and transfusion techniques.

\begin{figure}[t!]
\begin{center}

 \includegraphics[width=6.75in]{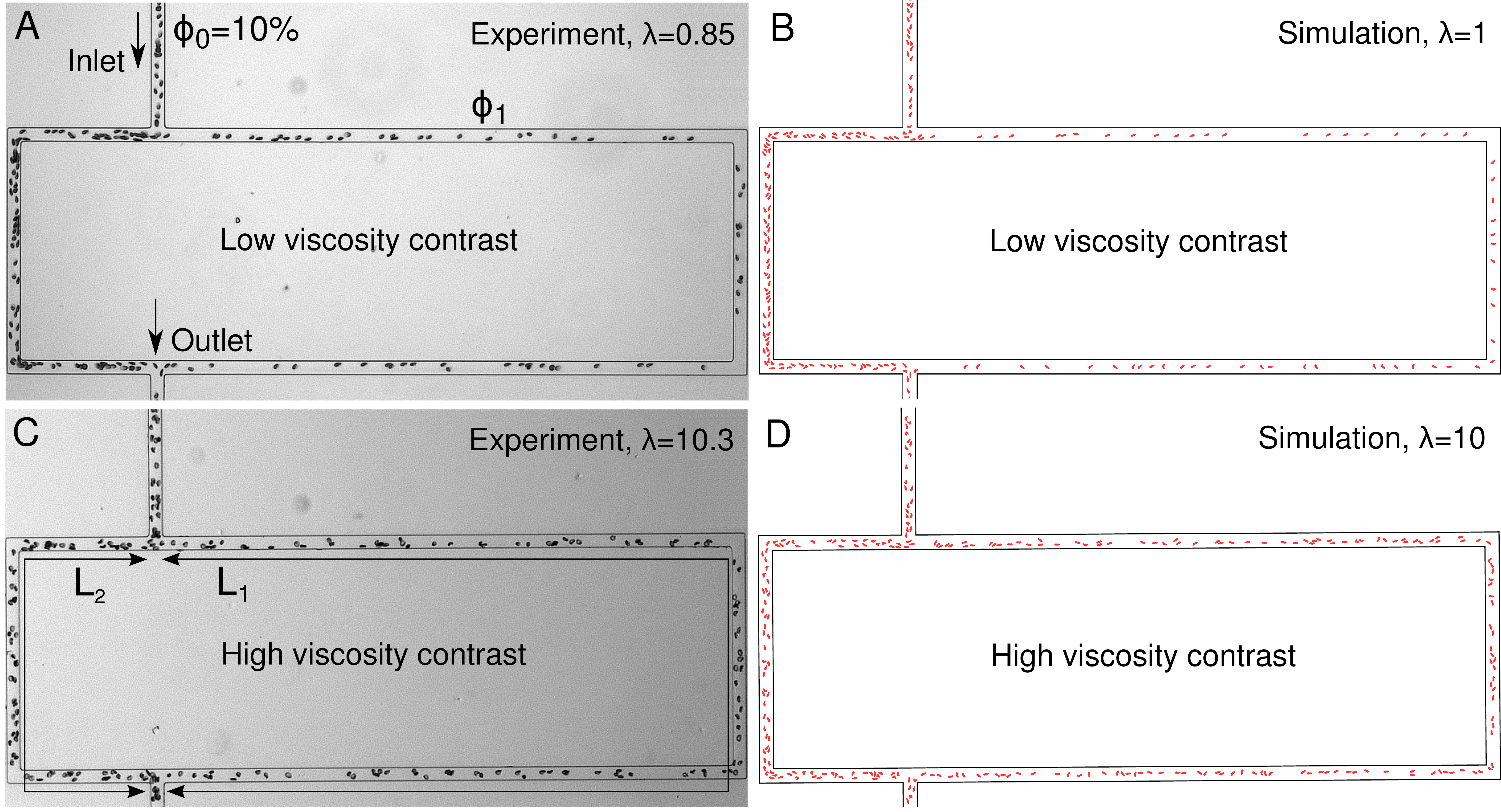}

 \caption{Snapshots of the RBCs partition, in both experiments and simulations, when the hematocrit of the feeding flow is around $\phi _0 =$10\%. The length ratio between the two child branches is set to 3. A, B: Low viscosity contrast (experiments with $\lambda$=0.85 and simulation with $\lambda$=1). C, D: High viscosity contrast  (experiments with $\lambda$=10.3 and simulation with $\lambda$=10).\label{figillus}}

\end{center}
\end{figure}

A major open problem in blood circulation is to understand the perfusion in the vasculature networks, especially in the microvasculature where RBCs accomplish their vital functions. For example, an improper hematocrit distribution is observed in heart microcirculation with consequences such as occlusion zones (within many patients with apparently healthy coronary arteries). These abnormal traffic zones cause a lack of oxygen supply to tissues that leads to cardiac ischemic disease \citep{cokkinos06}.
The precise origin of this dysfunction is still a matter of debate. The principal mechanism that dictates blood heterogeneity in the microvascular networks is the hematocrit partition at the level of bifurcations. RBCs do not behave as passive tracers. Their shape flexibility and dynamics have a decisive role because their size is comparable to that of blood capillaries. A well known phenomenon in microcirculation is the Zweifach-Fung effect \citep{dellimore83,fenton85,guibert10,pries89}: If we consider a bifurcation (as in Figure \ref{figillus}), the child branch with the lower flow rate is depleted in RBCs as compared to the parent vessel, while the other, higher flow rate child branch is enriched. That is, if in the parent vessel the total volumetric flow rate is $Q_0$ and the RBC volumetric flux is $N_0$, and in the child branch with the lower flow rate this flow rate is $Q_1$ and the RBC flux $N_1$, then $N_1/N_0<Q_1/Q_0$. When the flow rate is sufficiently small, the hematocrit in the child branch  can even drop down to zero, while it reaches high values in the other branch. Obviously, this phenomenon has physiological consequences as it alters the transport of oxygen and other essential metabolites, and may even trigger pathological disorders (e.g. occlusions in high hematocrit regions where the viscosity is higher and cell adhesion is favored).

Previous studies have investigated the impact of the confinement \citep{barber08,chien85,doyeux11,fenton85}, the bifurcation geometry \citep{audet87,hyakutake15,roberts03,roberts06,woolfenden11}, the hematocrit \citep{ditchfield96,fenton85,roberts03,yin13}, and the RBCs deformability \citep{barber08,li12,xiong12,yin13} and aggregation  \citep{sherwood14,yin13}. Most of these parameters influence RBCs distribution in the feeding flow, which is believed to dictates the partition at the bifurcation \citep{doyeux11,fenton85,li12,yin13}. The Zweifach-Fung effect results from the existence of a cell free layer (CFL) close to the walls, which is only occupied by plasma. The feeding flow is divided by a separating streamline into two parts, one feeding the low flow rate branch and the other feeding the high flow rate branch. Due to the CFL, the RBC fraction entering the low flow rate branch is smaller compared to the original RBC fraction in the total feeding flow. 
The depletion in the low flow rate branch is accompanied by enrichment in the high flow rate branch. In addition to the CFL as the main cause of the Zweifach-Fung effect, it has been argued that there is a relatively small counteracting effect where cells entering the bifurcation tend to be displaced towards the low flow rate branch compared to fluid streamlines (a fact that slightly reduces the Zweifach-Fung effect) \citep{barber08,doyeux11,li12,ollila13}, but this question is still  debated \citep{hyakutake15,xiong12}.

The existence of a CFL is a consequence of the lateral migration of RBCs towards the vessel center. This migration is a result of the wall-induced lift force due to hydrodynamic interactions \citep{callens08,cantat99,grandchamp13}, that depends on the nature of RBC dynamics (like tank-treading or tumbling \citep{abkarian07,dupire12}). 
The final distribution of RBCs in a confined straight vessel is not only due to the lift force, but it is also influenced by other factors: (i) The curved velocity profile of the Poiseuille flow \citep{coupier08,farutin14,katanov15,shi12}, and (ii) The cell-cell hydrodynamic interactions \citep{grandchamp13,hariprasad14,katanov15,kruger14,mcwhirter09}.

In the present work, we study the hematocrit partition at bifurcations using two-dimensional lattice Boltzmann simulations, whose outcomes are validated and supported by microfluidic experiments. We show that RBCs deformability  strongly impacts partition as long as the hematocrit is below $20\%$ (within the normal range in microcirculation). RBC deformability is governed by several parameters such as membrane stiffness (shear, dilatation and bending elastic moduli), swelling degree, membrane viscosity and the viscosity contrast between the hemoglobin and the suspending fluid. Here we choose to tune the deformability through the latter parameter, the viscosity contrast, that controls the RBC  dynamics (tank-treading, tumbling or swinging)  then all the migration mechanisms at the origin of the CFL. On the other hand, and more importantly, this study reveals that hematocrit partition can be completely reversed, that is the low flow rate child branch can be enriched in RBCs compared to the parent vessel. This newly reported effect is an outcome of a subtle RBCs structuration in the microcirculatory system. This highlights the importance of the notion of RBCs spatio-temporal organization as the main non-negligible ingredient  to further understand blood perfusion in the microvasculature.

\section{Materials and methods}
\label{sec:methods}
\subsection*{Design of the microfluidic bifurcations} 
In both simulations and experiments, we use T-shaped bifurcations such as shown in Fig.\ref{figillus}: A parent channel divides into two child branches with the same width, but with different lengths $L_1$ and $L_2$ ($L_1>L_2$). The ratio of the flow rates in branches 1 and 2 is then given by $Q_1/Q_2=(\eta_2 L_2)/(\eta_1 L_1)$, where $\eta_1$ and $\eta_2$ are the apparent viscosities of the suspension in branches 1 and 2, respectively. For dilute suspensions, where the viscosity is close to that of the suspending fluid, we simply have $Q_1/Q_2=L_2/L_1$. In simulations, we set the width of the channels to $W=20 \mu$m and we vary $L_1/L_2$ from 1.43 to 3. In experiments, we have $L_1/L_2=3$, $W=20\mu$m and the height of the channel $h$ is 8 $\mu$m. The length of the parent vessel was chosen as long as possible to allow for the development of a stationary distribution of RBCs across the channel in the feeding flow (5 mm in experiments and 1.5 mm in simulations). Microfluidic channels were produced by standard soft lithography techniques, with molded PDMS bonded to glass. The RBC suspensions were perfused by a syringe pump (KDS Legato 180) and imaging was performed by a video camera (Imaging Source DMK 31AF03) mounted on an inverted microscope with motorized stage (Olympus IX71) and a blue filter ($434 \pm 25$ nm) corresponding to an absorption peak of hemoglobin.

\subsection*{Blood preparation}
Blood samples were provided by the Etablissement Fran\c{c}ais du Sang (EFS Rh\^{o}ne-Alpes) from healthy donors. 
RBCs were isolated by centrifugation after being washed twice in phosphate buffer saline (PBS) supplemented by 0.1 \% bovine serum albumin (BSA). 
To prevent sedimentation of RBCs in channels, the RBCs were re-suspended in density matching PBS and BSA solutions in a mixture of water and iodixanol (Optiprep from Axis-Shield). 
This iso-dense solution was used either alone (1.94 mPa.s at 20 $^{\circ}$C) or after adding 5\% dextran of molecular weight $2\times10^6$ (viscosity 23.4 mPa.s at 20 $^{\circ}$C). 
The viscosity of the internal hemoglobin solution of healthy RBCs is around 20 mPa.s at 20 $^{\circ}$C \citep{kelemen01}. 
This provides two values of the viscosity contrast $\lambda$, namely  10.3 and 0.85. The first value corresponds to the blood at 20 $^{\circ}$C, while the physiological value at body temperature is around 5-6 \citep{cokelet68}. 

Note that we chose to vary the viscosity contrast $\lambda$ as one way to tune deformability, and therefore the
dynamics of lift and hydrodynamic interactions of cells. Stiffening cells using diamide or glutaraldehyde was another possibility. However, from the experimental viewpoint, working with hardened cells  at high
volume fractions in such a confined environment is quite difficult due to jamming. It would have been nearly impossible to
inject a suspension of very stiff cells at hematocrits larger than
10\%. Also, the dynamics of glutaraldehyde hardened cells is pure tumbling,
which corresponds to very high values of the viscosity ratio $\lambda$.
We do not expect the dynamics (and therefore phase separation) to
change much at values of $\lambda$ greater than 10 and we found more
interesting to increase deformability by decreasing $\lambda$ rather
than trying to investigate less deformable cells (with the
experimental difficulties mentioned above).

\subsection*{Hematocrit measurements} 
Local hematocrit measurements were made by comparing suspension flow images to a reference image without RBC, under identical illumination, and using the Beer-Lambert law of absorption. 
The absorption coefficient was determined by a calibration with images at low hematocrit, where a direct measurement can be made by counting individual cells. Hematocrit in branch $i$ will be denoted $H_i$.
In experiments, $h$ is small enough ($8 \mu$m) so that the flow is quasi two-dimensional. To allow a qualitative comparison with 2D numerical simulations, an area hematocrit $\phi_i$ was also derived by multiplying the number of cells per unit area by the average cross-sectional area of RBCs ($S=$19.8 $\mu$m$^{2}$). $H_i$ and $\phi_i$ are therefore linked by the relationship $H_i=\phi_i v/(S h)$, where $v=90$ $\mu$m$^{3}$ is the average volume of one cell.

 \subsection*{Simulation method} 
In simulations, we use lattice Boltzmann method (LBM) to compute the fluid flow \citep{kaoui11,zhang07}. 
Each RBC is modeled by 60 nodes interconnected by a potential that allows bending, as well as a stretching modulus that penalizes distance variations between two adjacent nodes. 
This achieves the RBC membrane incompressibility \citep{tsubota06}. 
In other words, we set the spring constant to values as large as possible in order to keep the ratio between the membrane perimeter and area constant. 
We define the reduced area as $4{\pi}A/p^2$ (with p is the perimeter and A the enclosed area), which we set to 0.7 to produce a RBCs with a biconcave shape. 
We use the immersed boundary method (IBM) to couple the fluid flow and RBC deformation \citep{kaoui11,peskin02,zhang07}. 
For comparison with experiments, we set the viscosity contrast to $\lambda=1$ and $\lambda=10$.

\begin{figure}[t!]
\begin{center}

 \includegraphics[width=3.25in]{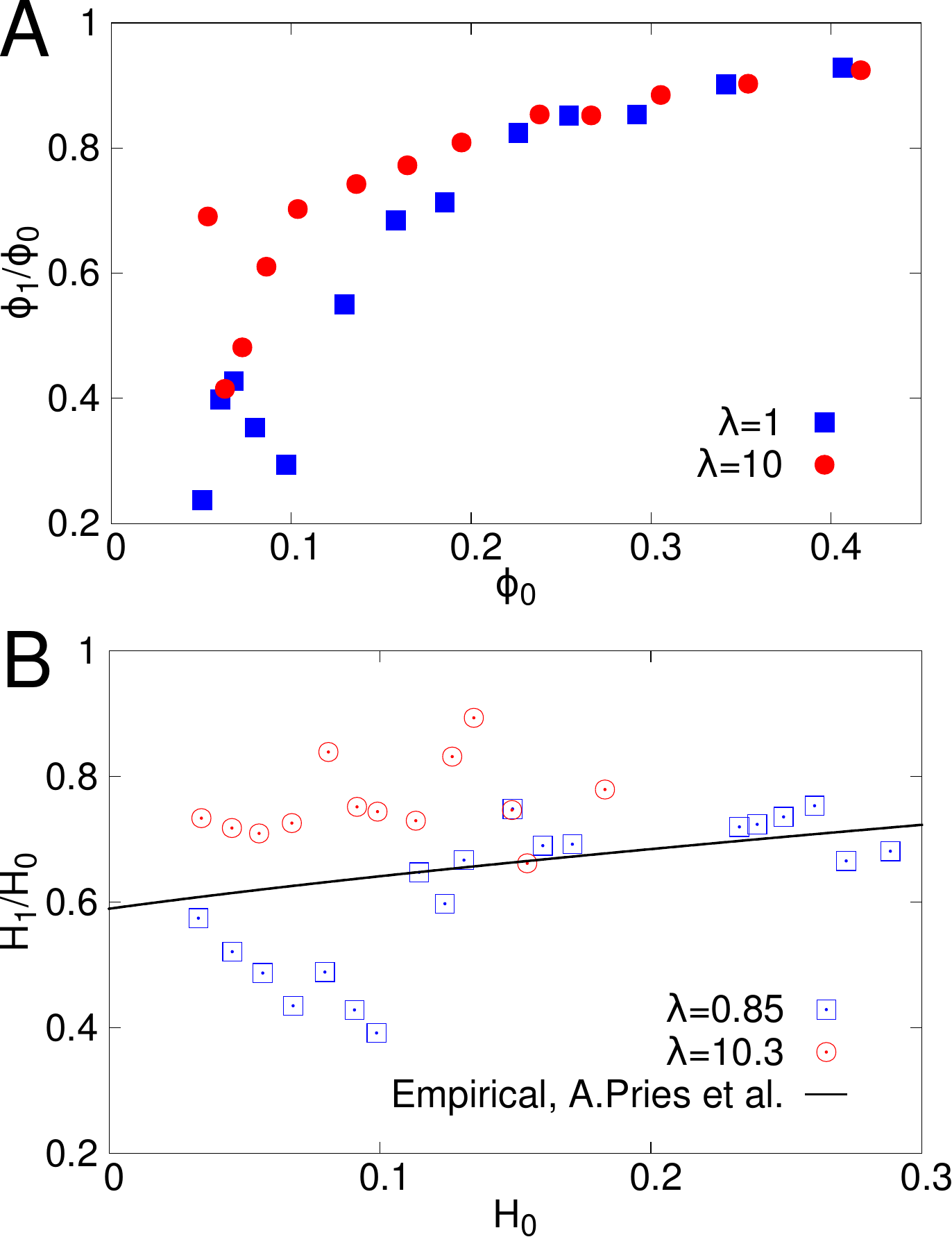}

 \caption{The relative hematocrit in the low flow rate branch as a function of the hematocrit in the parent vessel. The length ratio between the branches  $L_1/L_2$ is set to 3. A: Simulations with $\lambda$=1 and $\lambda$=10. B: Comparison between experiments ($\lambda$=0.85 and $\lambda$=10.3) and the empirical law of Pries \textit{et al.} \citep{pries90}  (solid line), for the same cross-sectional area. The non-monotonous evolution of the relative hematocrit at low $H_0$ and high $\lambda$ is related to the inversion of the Zweifach-Fung effect, on which we comment later on (see Fig.  \ref{fig-antiZFexp}).\label{figphi1phi0}}

\end{center}
\end{figure}

\section{Results}
\label{sec:results}

As a guideline, we shall refer to the empirical law of Pries \textit{et al.} taken from refs \citep{pries89,pries90} that gives the hematocrit partition at a bifurcation: 
 \begin{equation}
\mathrm{logit}\big(\frac{H_1Q_1}{H_0 Q_0 }\big)=\alpha \mathrm{logit}\big(\frac{Q_1/Q_0-\beta}{1-2 \beta}\big),\label{eqpart}
 \end{equation}
 where $\alpha=1+ 6.98 (1-H_{0_F})/a$, $\beta=0.4/a$ (with $a$  the tube diameter in microns) and $\mathrm{logit}(x)=\ln[x/(1-x)]$. $H_0$ is the volumetric hematocrit in the parent feeding branch, while $H_1$ is the hematocrit in a child branch. $H_{0_F}$ is the feeding hematocrit in a reservoir that would be located right before the narrow feeding vessel. Due to the F\aa hr\ae us effect, $H_{0_F}$ is larger than $H_0$ and a relationship between both quantities is also given in \cite{pries90}: 
 \begin{equation}
 H_0/H_{0_F}=H_{0_F}+(1-H_{0_F})(1+1.7 e^{-0.415 a}-0.6 e^{-0.011 a}).
 \end{equation}
Note that the partition law (Eq. \ref{eqpart}) has been validated through \textit{in-vivo} experiments with rats (thus at body temperature), with narrow capillaries (of diameters $a$ lower than 30 microns), but with feeding hematocrits higher than 20\%. 

As we shall compare predictions for 3D hematocrits in a cylindrical tube with either 2D simulations or experiments in a rectangular channel, we should avoid any direct quantitative comparisons, but rather use Pries \textit{et al.} predictions as a guideline to identify where new behavior is exhibited. For comparison with simulations, we set $a=W$, where $W$ is the channel width, and we shall  consider only the  hematocrit ratios. For the experiments in rectangular channels, we set $a$ to adjust the cross-sectional areas: $\pi a^2/4=Wh$.

\subsection*{The role of interactions}

We analyze in details how RBC deformability affects the hematocrit partition at the bifurcations. 
Figure \ref{figillus} illustrates the Zweifach-Fung effect, observed in both experiments and simulations, at a feeding area hematocrit of 10\%. 
In both cases, less RBCs enter the low flow rate branch (the long branch) simply due to the flow rates differences between the two child branches. 
However, the asymmetry is significantly pronounced at low viscosity contrast $\lambda$ (when the suspending fluid is more viscous than the hemoglobin). 
To quantify the partition asymmetry, we measure the relative hematocrit $\phi_1/\phi_0$ (or, equivalently,  $H_1/H_0$), in the low flow rate branch, while we vary the hematocrit in the parent branch (Fig. \ref{figphi1phi0}A).  Either in the simulations (Fig. \ref{figphi1phi0}A) or in the experiments (Fig. \ref{figphi1phi0}B), we see less RBCs in the low flow rate branch than in the parent one ($\phi_1/\phi_0<1$), when the inlet hematocrits ($\phi_0$ or $H_0$) lies between $5\%$ and $45\%$,  which is precisely a manifestation of the  Zweifach-Fung effect.
When the viscosity contrast is low, we observe a significantly strong reduction of hematocrit in the low flow rate branch, both in experiments and simulations, at moderate inlet hematocrit. This interesting observation suggests that the RBCs mechanical properties can strongly impact the hematocrit partition \textit{in-vivo} since the normal hematocrit is usually less than $20\%$ (typically between 10 and 20 $\%$ \citep{Fung13}) in microcirculation.

\begin{figure}[t!]
\begin{center}

 \includegraphics[width=6.25in]{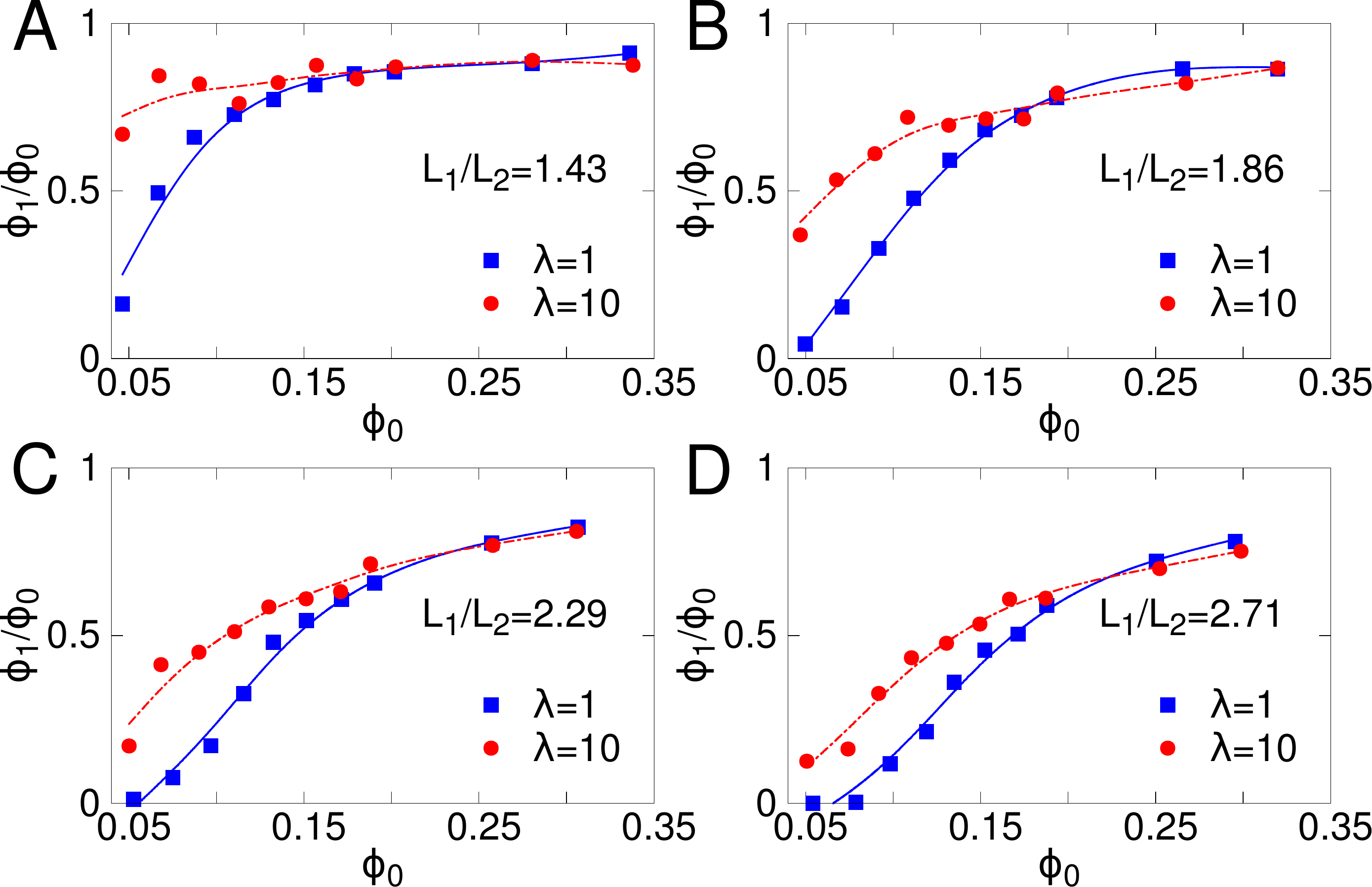}

\caption{Simulations: the relative hematocrit in low flow rate branch $\phi_1/\phi_0$ as a function of the hematocrit in the parent vessel $\phi_0$, for several branches length ratios $L_1/L_2$ and viscosity contrasts $\lambda$.  At low enough $\phi_0$, the asymmetry between the two daughter branches is strongly enhanced as the viscosity contrast $\lambda$ is decreased, while the partitioning becomes independent on $\lambda$ for hematocrits above 20$\%$. \label{figdepL1L2}}

\end{center}
\end{figure}

However, when the hematocrit is high enough,  the viscosity contrast plays a minor role. This is clear in simulations (for $\phi_0$ larger than 25\%, Fig.  \ref{figphi1phi0}1). Similarly, in the experiments, above $H_0=20\%$, data for both $\lambda$ converge to the Pries \textit{et al.} prediction. The insensitivity to the viscosity contrast beyond a critical hematocrit ($\phi_0\simeq 25\%$) is a robust phenomenon that is independent of the length ratio between the branches (i.e. roughly the bulk flow rate ratio), as illustrated on Figure \ref{figdepL1L2}).

\begin{figure}[t!]
\begin{center}

 \includegraphics[width=6.25in]{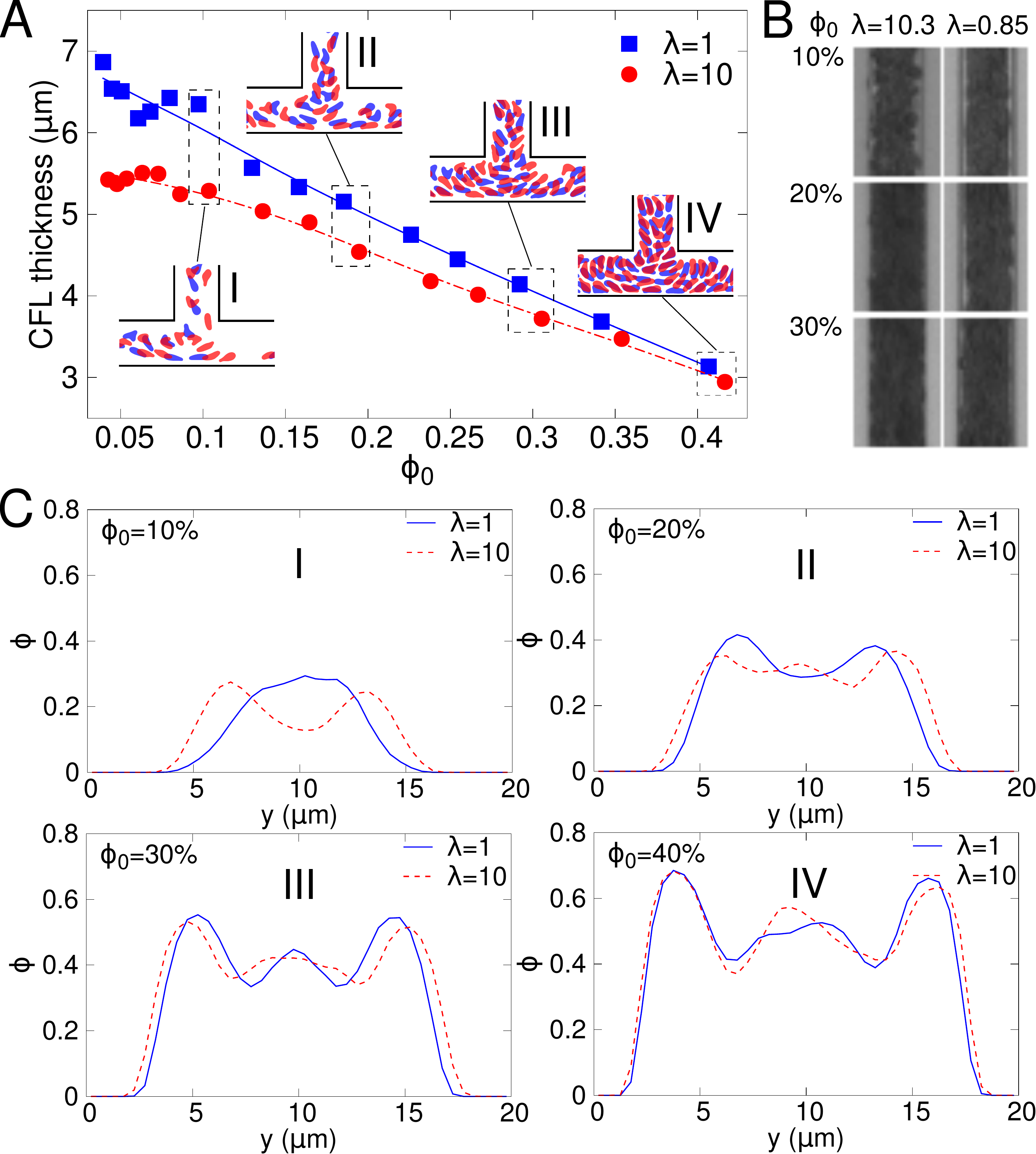}

 \caption{A: CFL thickness as a function of the hematocrit in the parent vessel, for $\lambda$=1 and $\lambda$=10, in simulations. Insets : Snapshots showing the suspension at the bifurcation. We define the CFL as a layer where the integrated concentration profile  is below $5\%$ \citep{kumar14}. B: Snapshots from experiments, for $\lambda$=0.85 and 10.3, and hematocrits $\phi_0=$10, 20 and 30\%. Every snapshot is a superimposition of 10 successive images in order to highlight the CFL in the parent vessel. C: The stationary volume fraction density functions in the parent vessel obtained from simulations.\label{figprofiles}}

\end{center}
\end{figure}

It is appealing to suggest that the dependency of the hematocrit partition upon the feeding hematocrit is the result of the up-stream organization of RBCs in the parent vessel due to hydrodynamic interactions. 
At low hematocrit flows, the cell-cell interaction is weak and the organization of RBCs, within the vessel, depends mainly on the dynamics of each RBCs, thus on $\lambda$.
The RBCs aggregate at the center of the vessel due to the wall-induced lift force, that increases with decreasing $\lambda$ and increasing RBCs deformability \citep{grandchamp13}. 
This means that suspensions of RBCs with high viscosity contrasts have wider distributions (smaller CFL) in the channel as compared to suspensions having lower viscosity contrast.
As a consequence, the asymmetry in the partition is expected to increase when the viscosity contrast decreases, as shown in Figure \ref{figdepL1L2}. 
To support this argument, the CFL thickness in the parent vessel and the configuration of RBCs before the bifurcation are reported in Figure \ref{figprofiles}A,B.
We can clearly see that RBCs distribution at high viscosity contrast ($\lambda$=10 in simulations and $\lambda$=10.3 in experiments) is wider than that at low viscosity contrast ($\lambda$=1 in simulations and $\lambda$=0.85 in experiments) when the feeding hematocrit is low ($\phi_0<$20\%, see also Figure \ref{figprofiles}C-I, II).

However, when the hematocrit increases, the contribution of the hydrodynamic interactions among RBCs becomes stronger and stronger. This causes a broadening of the distribution that acts against the lift force. 
Consequently, the partition between the two branches becomes more symmetric (that is $\phi_1/\phi_0$ becomes close to 1). Interestingly, those broad distributions are quasi independent of the viscosity contrast (see  Figure \ref{figprofiles}A and C-III, IV). Consequently, $\phi_1/\phi_0$ does not depend on $\lambda$ either (see Figure \ref{figdepL1L2}). 

Thus the distribution is independent of the strength of the interactions between cells and between cells and walls, but it is mainly caused by geometrical constraints. 
In other words, interaction between cells and the lift forces both depend on $\lambda$, and this result indicates that they depend more or less on $\lambda$ in the same way. The $\lambda$ contributions cancel out once a critical feeding hematocrit is reached. 
Noteworthy, beyond this critical hematocrit the separating ratio $\phi_1/\phi_0$  quasi plateaus which enforces the idea that in this regime, the feeding flow can be considered as a three-layer fluid (fluid-cell-fluid).
The width of each layer will depend neither on the strength of interaction (which is related to deformability) nor on the volume fraction.

 \begin{figure}[t!]
\begin{center}

 \includegraphics[width=3.25in]{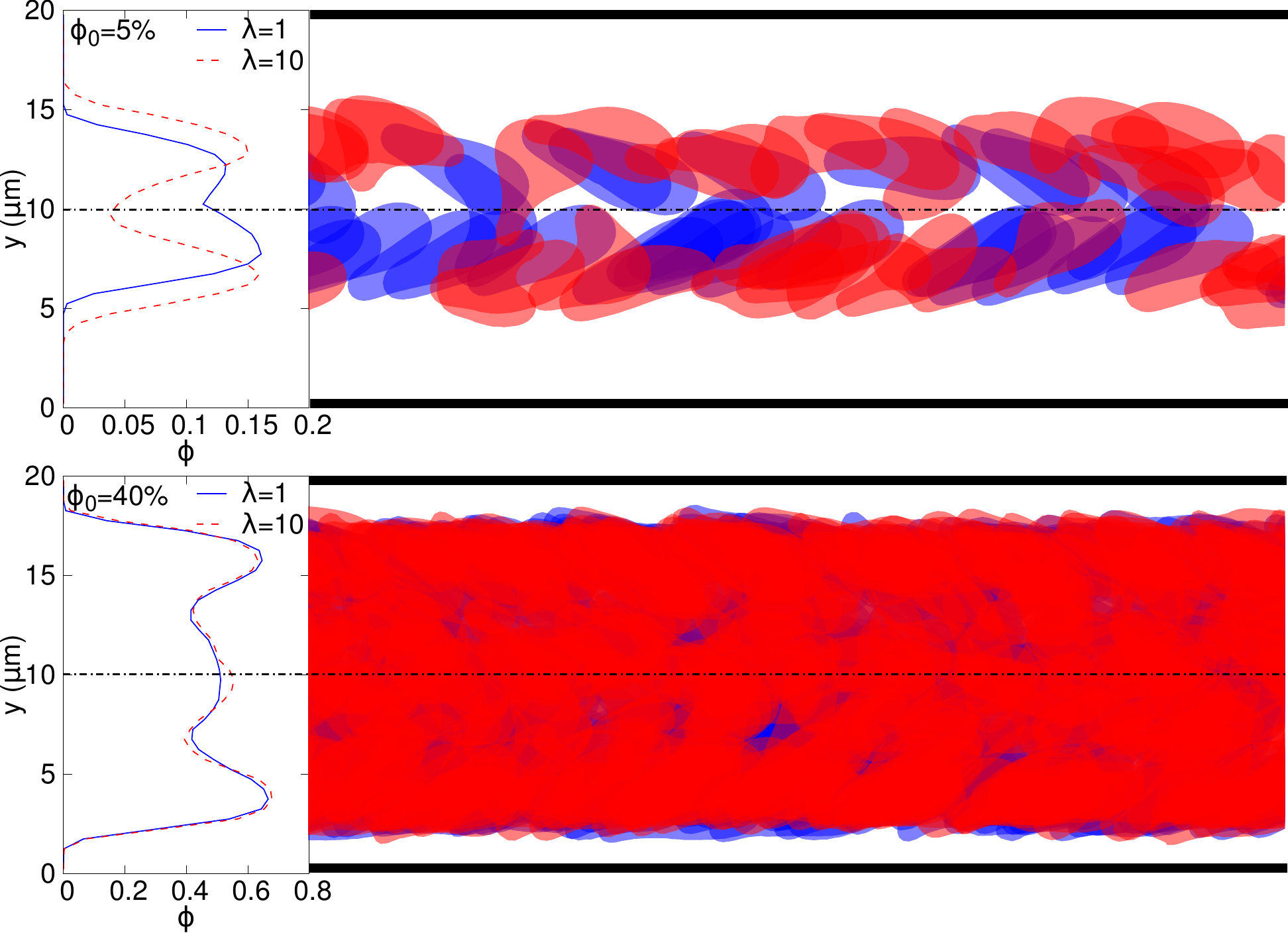}

 \caption{The profiles of the hematocrit distribution and the corresponding snapshots of RBCs distribution in the parent vessel. The feeding hematocrits are 5\% and 40\%, for two different values of $\lambda$. \label{fig5-40}}

\end{center}
\end{figure}

\subsection*{The low hematocrit limit: Inversion of the Zweifach-Fung effect}

\begin{figure}[t!]
\begin{center}
 \includegraphics[width=3.25in]{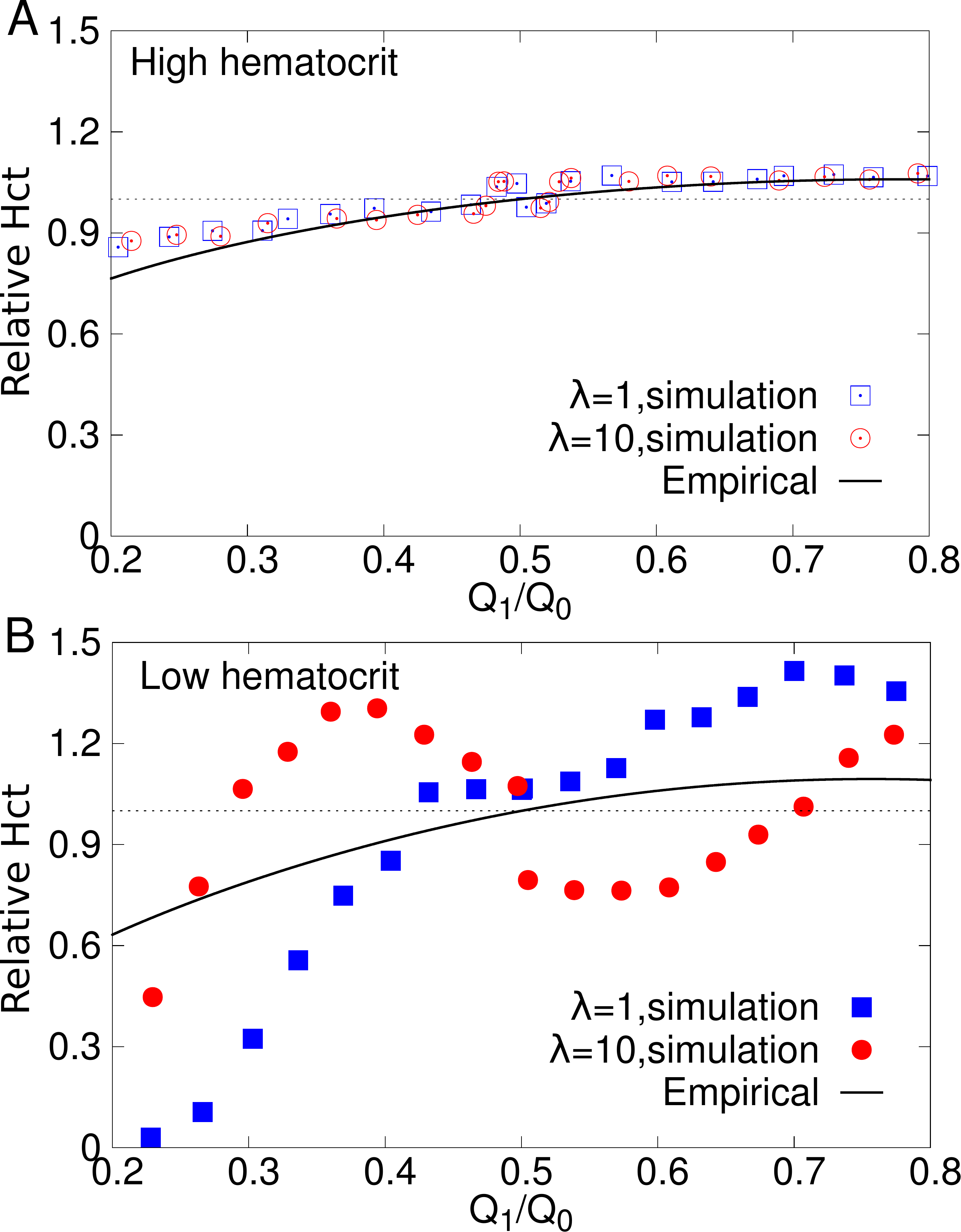}

\caption{The relative hematocrit in one child branch as a function of the bulk flow ratio. Solid lines correspond to the empirical law proposed in ref \citep{pries90}, for $a=W$. For simulations, the relative hematocrit is $\phi_1/\phi_0$. For Pries law, it is given by $H_1/H_0$. A: high hematocrit ($\phi_0=H_0=40\%$). B: low hematocrit ($\phi_0=H_0=5\%$).\label{fig-antiZF}}
\end{center}
\end{figure}

Now we focus on the low hematocrit case, for which the partition depends strongly on the detail of the interactions and on the volume fraction. Moreover a peculiar effect arises due to the prevalence of the discrete nature of blood at that scale. 
For all hematocrits, the distribution of RBCs is not homogeneous, but rather exhibits two lateral peaks (Figure \ref{figprofiles}). 
This become more pronounced at low hematocrit ($\phi_0\lesssim 5 \%$), where a two-file distribution of RBCs is observed, as shown in Figure \ref{fig5-40}. 
For $\lambda$=10, there is almost no cell flowing in the central part of the vessel, even though the wall-lift force tends to center them. 
The structure adopted by the suspension can be viewed as a juxtaposition of layers with high and low hematocrits. 
For example, as shown in Figure \ref{fig5-40} (top panel with $\lambda=10$) the central part is depleted in RBCs, but it is escorted by two enriched layers, which themselves are surrounded by two depleted layers at the periphery (close to the channel walls). 
This 5-layer configuration (fluid-cell-fluid-cell-fluid) has an extremely interesting impact on the partition.
This can be highlighted by measuring $\phi_1/\phi_0$ as a function of the bulk flow rate ratio $Q_1/Q_0$ between a child branch and the parent vessel, for fixed $\phi_0$ (Figure \ref{fig-antiZF}).

If we focus first on the results for high $\phi_0$ ($\phi_0\simeq 40\%$), we find again the insensitivity to $\lambda$. 
As $Q_1/Q_0$ is increased from 0 to 0.5, the low flow rate branch 1 recruits first the CFL and then the cells.
This implies that $\phi_1$ starts at 0 and increases until reaching $\phi_0$ when the situation is symmetric ($Q_1=Q_2=Q_0/2$). 
Our results agree with  a previous 2D simulation obtained for $\phi_0\simeq32\%$ \citep{yin13} as well as with the empirical law of Pries and coworkers \citep{pries90}.


\begin{figure}[t!]
\begin{center}

 \includegraphics[width=6.25in]{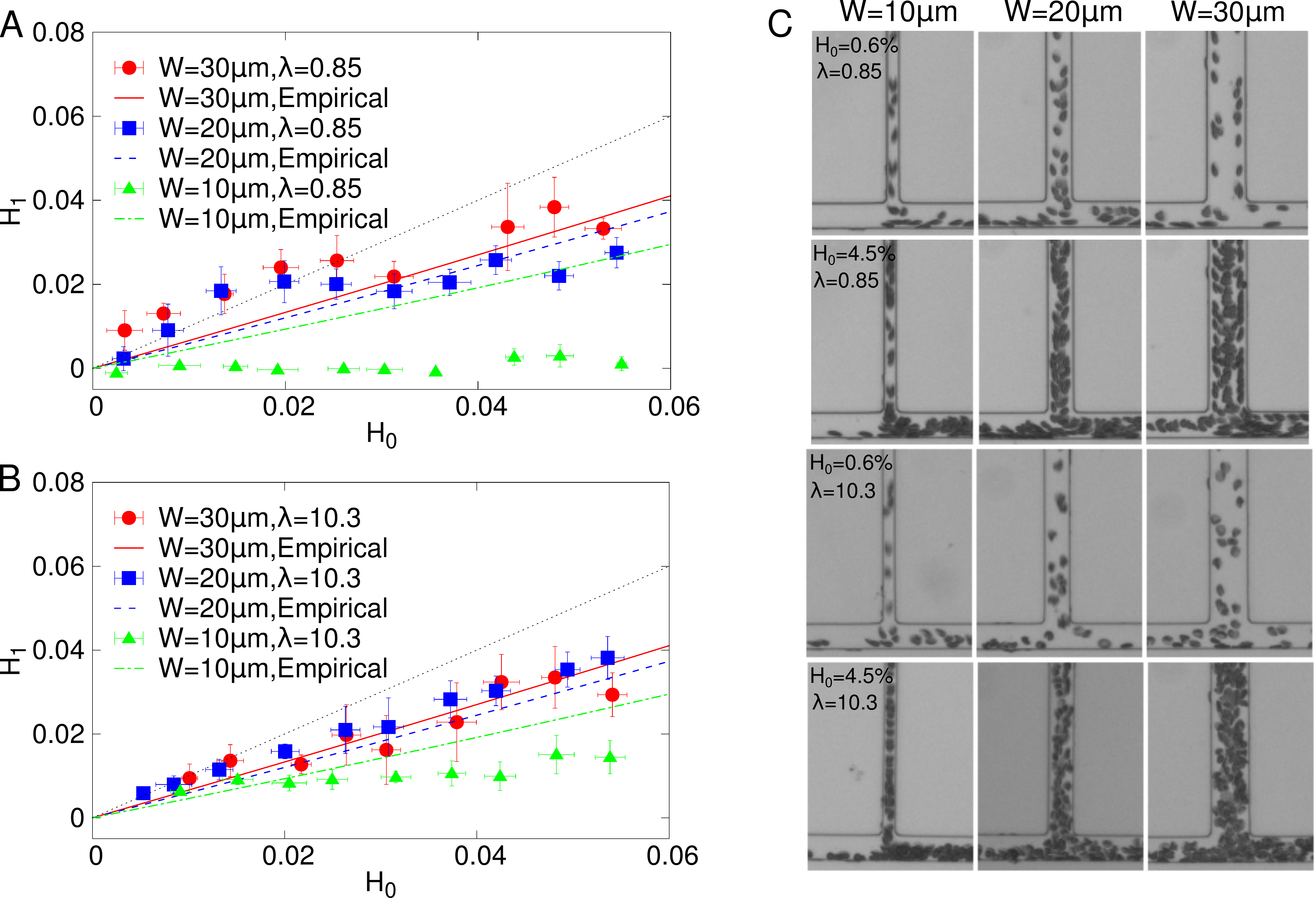}

 \caption{Experiments: the hematocrit in the low flow rate branch $H_1$ as a function of the hematocrit in the parent vessel $H_0$. A: Low $\lambda$ ; B: High $\lambda$. The length ratio between the branches is set to 3. The width $W$ of the inlet channel is set to 10, 20 and 30 $\mu$m. The corresponding empirical laws of Pries \textit{et al.}\citep{pries90} are also shown. The dotted line (the one with highest slope) corresponds to equal partition ($H_1=H_0$). For $W=20$ $\mu$m, the data are the continuation of the data already reported in Figure \ref{figphi1phi0}.C: Snapshots for each width $W$ and two concentrations $H_0=0.6\%$ and 4.5\%. Every snapshot is the superimposition of 10 successive images in order  to highlight the structure of the suspension in the parent vessel\label{fig-antiZFexp}}
\end{center}
\end{figure}

An unexpected phenomenon is observed at low enough hematocrit, for $\phi_0=5\%$ and at high viscosity contrast $\lambda=10$,
in contrast to the high hematocrit regime (see Figure \ref{fig-antiZF} B, $\lambda=10$).
At low $Q_1/Q_0$, the peripheral CFL in the parent vessel is recruited by the branch 1 so $\phi_1$ starts at 0 and increases when $Q_1$ increases. 
Around $Q_1/Q_0=0.3$, $\phi_1$ becomes larger than $\phi_0$.
This means the hematocrit is increased in the low flow rate branch, which is the reverse behavior of the Zweifach-Fung effect. 
The five-layer structure mentioned above is the key ingredient for understanding this unexpected behavior: 
In the intermediate range $0.3<Q_1/Q_0<0.5$, the low flow rate branch recruits the lateral CFL layer plus the adjacent RBC-rich layer among the five layers. 
By contrast, the high flow rate branch recruits the CFL layer close to the opposite wall plus its adjacent  RBC-rich layer (exactly as the low flow rate branch) as well as the central (and depleted) layer. 
Thus, while both branches recruit approximately the same amount of cells per unit time, those are more dilute in the high flow rate branch, which receives more fluid, while in the classical Zweifach-Fung effect, the high flow rate branch is the one that receives more cells. For $\lambda=1$ (see Figure \ref{fig-antiZF}B with $\lambda=1$), the two-peak structure is not as marked as in the case of $\lambda=10$, so the reverse Zweifach-Fung effect is not as strong. 
The subtle role played by the suspension structuring at low hematocrit is also supported by our experiments, where the interplay between the diffusion and the wall-lift force is controlled by varying the width $W$ of the inlet channel (see Figure \ref{fig-antiZFexp}). 
When $W$ is low ($W$ = 10 $\mu$m), the hematocrit in the low flow rate branch is much lower than expected from Pries \textit{et al.} predictions (which were not validated on this confinement range).
This is caused by the CFL effect that becomes very strong. 
As in the simulations, for $Q_1/Q_0=0.25$, this effect is more pronounced at low $\lambda$, that corresponds to a more important wall lift force. 
The 5-layer structure is clearly observed also in the experiments for $H_0<5\%$, $W=20$ or 30 $\mu$m and at low $\lambda$ (Figure \ref{fig-antiZFexp}C), but not at high $\lambda$, while it was more strongly marked at high $\lambda$ in the simulations. 
This indicates that this peculiar structure is very dependent on the mechanical properties of the cells and also on the degree of confinement. 
Nevertheless, a robust feature is valid in both simulations and experiments: When the two-file structure of RBCs takes place, a clear inversion of the blood partition at the bifurcation is observed. In the experiments, this corresponds to Figure \ref{fig-antiZFexp}A, where some points lie above the equal partition line when $H_0<2\%$.

There are also situations in which one of the two branches can be even completely devoid of RBCs (Figure \ref{fig-antiZFexp}C). A corresponding prolonged lack of RBCs perfusion to real blood vessels causes dysfunction and possibly ischemia disease. 
Because RBC mechanical properties are affected by aging and pathologies, these can induce abnormal partitions of the hematocrit in the vascular network. 

\section{Discussion and conclusions} 
\label{sec:discussion}

As a result of the interplay between the Zweifach-Fung effect and the F\aa hr\ae us effect the hematocrit in microcirculation can reach values as low as 10-20\% compared to the average hematocrit in human body (45\%).
At such a low hematocrit, our simulations and \textit{in vitro} microfluidic experiments have revealed that RBCs partition at the level of bifurcations depends strongly on the viscosity contrast between the viscosities of the RBC hemoglobin and the suspending fluid.
In the extreme hemodilution, our results exhibit a newly reported phenomenon: The low flow rate branch may receive higher hematocrit than the high flow rate branch in opposition to the known Zweifach-Fung effect. This phenomenon is observed under moderate confinement and is the result of a peculiar structuring of the cell suspension. Under stronger confinement, other strong discrepancy with Pries \textit{et al.} empirical law  was highlighted, with  a strong asymmetry in the partitioning (enhanced Zweifach-Fung effect). Our findings suggest that the viscosity contrast must be taken into consideration and carefully analyzed in order to have a firm understanding of RBC distribution in microcirculation.  This physiological parameter increases with aging as well as with some pathologies.

The results of our present work provide a valuable background needed to pinpoint the various RBCs properties that govern hematocrit partition, and thus oxygen delivery in the microcirculation in general. 

 
\section*{Acknowledgment}
Z. S. thanks Vassanti Audemar for experimental advice and assistance. Z. S., G. C., C. M. and T. P. acknowledge financial support from CNES (Centre National d'Etudes Spatiales) and ESA (European Space Agency). The DyFCom team of LIPhy (Z. S., G. C., C. M. and T. P.) is member of Labex TEC21 (Investissements d'Avenir-Grant Agreement ANR-11-LABX-0030), Structure F\'ed\'erative de Recherche iDYSCO (CNRS), F\'ed\'eration Galileo Galilei de Grenoble (FR3345 CNRS-UJF-Grenoble INP-IRSTEA) and Groupement de Recherche MECABIO (GDR3570 CNRS).


\end{document}